\documentclass[10pt,a4paper]{article}

\usepackage{jheppub}

\usepackage{epsfig,epsf}
\usepackage{amsmath}
\usepackage{amsthm}
\usepackage{amsfonts}
\usepackage{amssymb}
\usepackage{dsfont}
\usepackage{epstopdf}
\usepackage{multirow}
\usepackage{marvosym}
\usepackage{etoolbox}
\apptocmd{\thebibliography}{\setlength{\itemsep}{3.5pt}}{}{}

\usepackage{slashed}
\usepackage{float}

\usepackage[active]{srcltx}

\def\II{\hbox{{1}\kern-.25em\hbox{l}}}

\def\II{\hbox{{1}\kern-.25em\hbox{l}}}

\title{
\begin{flushright}
{\large \textnormal{DESY 17 -- 095}}\\[2mm]
{\large \textnormal{LMU-ASC 36/17}}\\[2mm]
\end{flushright}
\vspace{1cm}
Higher spin currents in the critical $O(N)$ vector model at~$1/N^2$}
\author[a,b]{A. N. Manashov,}
\author[c,d]{ E. D.  Skvortsov}
\author[b]{and \  M.   Strohmaier}

\affiliation[a]{
   Institut f\"ur Theoretische Physik, Universit\"at Hamburg,   D-22761 Hamburg, Germany}

\affiliation[b]{
   Institut f\"ur Theoretische Physik, Universit\"at
   Regensburg, D-93040 Regensburg, Germany}

\affiliation[c]{Arnold Sommerfeld Center for Theoretical Physics, Ludwig-Maximilians University Munich, \\
Theresienstr. 37, D-80333 Munich, Germany}

\affiliation[d]{Lebedev Institute of Physics,
Leninsky ave. 53, 119991 Moscow, Russia}

\emailAdd{alexander.manashov@desy.de}
\emailAdd{evgeny.skvortsov@physik.uni-muenchen.de}
\emailAdd{matthias.strohmaier@ur.de}

\abstract{We calculate the anomalous dimensions of higher spin singlet currents in the critical $O(N)$ vector model at order $1/N^2$. The results are shown to be in agreement with the four-loop perturbative computation in $\phi^4$ theory in $4-2\epsilon$ dimensions. It is known that the order $1/N$ anomalous dimensions of higher-spin currents happen to be the same in the Gross-Neveu and the critical vector model. On the contrary, the order $1/N^2$ corrections are different. The results can also be interpreted as a prediction for the two-loop computation in the dual higher-spin gravity.  }


\begin{document}

\maketitle

\section{Introduction}\label{Sec:Introduction}
The $N$-component  $\varphi^4$ model possesses a nontrivial critical point in $2<d<4$ dimensions and serves as an example
of a conformal field theory (CFT), see e.g. Ref.~\cite{ZinnJustin:1989mi}. The fundamental renormalization group functions
in this model 
are known with a high precision in the perturbative
expansion~\cite{Brezin73,Kazakov:1979ik,Dittes:1977aq,Chetyrkin:1981jq,Gorishnii:1983gp,Kleinert:1991rg,Kompaniets:2017yct}
that allows one to get  reliable predictions  for the critical indices in the physically interesting dimension $d=3$. This
model can also be analyzed within  the $1/N$ expansion framework.  This technique is very suitable for the description of phase
transition phenomena~\cite{ZinnJustin:1989mi}. Critical indices in this approach are given by a series in
$1/N$ with the coefficients being  functions of the space  dimension~$d$ that allows one to obtain indices directly in $d=3$.
Moreover, consistency of  the results of the perturbative  and $1/N$ expansions provides a nontrivial check of the calculations in
both approaches. Unfortunately, the calculations in the $1/N$ approach are rather involved. Only two indices~--~the critical
dimension of the basic field in the $N$-vector and Gross-Neveu models~--~are known to $1/N^3$
accuracy~\cite{Vasiliev:1982dc,Vasiliev:1992wr,Gracey:1993kc}. Nevertheless  many critical exponents are available at the $1/N^2$
order, see for a review~\cite{Vasilev:2004yr}.

Recent interest in the $O(N)$-vector model comes from studies of the AdS/CFT correspondence. Namely, it was conjectured in
\cite{Klebanov:2002ja} that the critical $O(N)$ vector model should be dual to the higher-spin theory in AdS${}_4$, see also
\cite{Sezgin:2002rt,Sezgin:2003pt,Leigh:2003gk}. Some tests of this conjecture have been already performed at the level of
tree-level three-point correlation functions and one-loop determinants~\cite{Giombi:2009wh,Giombi:2013fka,Giombi:2014yra}. Although
the conjectured duality is supported by  these tests it goes without saying that verification  beyond the tree level is desirable.
The simplest quantities  to compare on both sides of this correspondence are the masses (AdS) and anomalous dimensions (CFT) of the
currents. Indeed, it is expected that the radiative corrections on the AdS side should give masses\footnote{ The masses are
measured in the units of the cosmological constant. } to  higher-spin fields~\cite{Girardello:2002pp,%
Manvelyan:2008ks,Skvortsov:2015pea}, that, from the CFT point of view, correspond to the anomalous dimensions,
$\gamma_s$, of the higher-spin currents
\begin{align}
m_s^2=m_0^2(s)+\delta m_s^2\,, && m_0^2(s)=(d+s-2)(s-2)-s,&&  \delta m_s^2=\gamma_s(d-4+2s+\gamma_s)\,.
\end{align}
The anomalous dimensions are known at the $1/N$ order for the singlet  currents  ~\cite{Lang:1991kp} and at $1/N^2$ for the
non-singlet currents~\cite{Derkachov:1997ch}. The aim of this work is to bridge this gap and calculate the anomalous dimensions
of the singlet currents to the $1/N^2$ accuracy.

The higher-spin vs. vector model duality turns out to be a particular case of a more general duality between Chern-Simons
matter theories and parity breaking higher-spin theories, \cite{Giombi:2011kc,Aharony:2011jz}. There are four simplest
Chern-Simons matter theories: free boson, critical vector model, free fermion and Gross-Neveu models that are coupled to a
Chern-Simons gauge field at level $k$. The three-dimensional bosonization duality \cite{Giombi:2011kc, Maldacena:2012sf,
Aharony:2012nh} identifies these four theories pairwise. The AdS/CFT duality then relates them to a parity-violating
higher-spin theory in $AdS_4$ with two different boundary conditions for the scalar field of the higher-spin multiplet.
In this more general picture the Gross-Neveu model and the critical vector model turn out to be duals of one and the same higher-spin
theory, but for different values of the parity violating parameter and boundary conditions. Remarkably, the anomalous
dimensions of the singlet higher-spin currents happen to be the same at order $1/N$ in $d=3$ both for the critical vector
model and the Gross-Neveu model \cite{Muta:1976js}:
\begin{align}
\gamma_{s}&= \frac1{N}\frac{16 (s-2)}{3 \pi ^2 (2 s-1)}\,\qquad \text{for even } s\,.\label{GNWF}
\end{align}
Recently, the order-$1/N$ anomalous dimensions have been computed in all the four basic Chern-Simons matter theories
\cite{Giombi:2016zwa}. The result is that they are given by two functions of spin, one of them being $\gamma_{s}$ above, times
simple factors that depend on the parity violating parameter. It is interesting that the two spin-dependent functions were
found to be same for all the four theories, a particular case being \eqref{GNWF}. In \cite{Manashov:2016uam} the order-$1/N^2$
anomalous dimensions were computed for the Gross-Neveu model. The results of the present paper reveal that the order-$1/N^2$
anomalous dimensions are different in the critical vector and the Gross-Neveu models. It would be interesting to extend the results to Chern-Simons matter theories.

The paper is organized as follows. In Section~\ref{sect:model} we describe the model and review a technique for the calculation
of critical exponents. In Section~\ref{sect:HSO} we discuss the renormalization of higher-spin currents and present our
results for the anomalous dimensions of these currents in the $1/N^2$ order in arbitrary dimension $d$. The anomalous dimensions
in
$d=3$ are discussed in Section~\ref{sect:d3}. The details of the calculations and some numerical results are collected in several
Appendices.

\section{Critical $O(N)$ model}\label{sect:model}
The  $O(N)$ invariant $\varphi^4$ model (where $\varphi$ is an $N$-component real field) 
%
\begin{align}\label{Sphi4}
S(\varphi)=\int d^d x \left\{\frac12 (\partial\varphi)^2+\frac{1}{4!}g M^{2\epsilon}(\varphi^2)^2\right\}\,,
\end{align}
%
has a nontrivial Fisher-Wilson
critical point
 in $d=4-2\epsilon$ dimensions~\cite{Wilson:1971dc,Wilson:1973jj},
\begin{align}\label{critical-u}
u_* &=\frac{6\epsilon}{N+8}\biggl(1+\frac{6\epsilon(3N+14)}{(N+8)^2}
\notag\\
&\quad +\frac{\epsilon^2}{2(N+8)^4}
\Big(-33N^3+110N^2+1760N+4544-96\zeta_3(N+8)(5N+22)\Big)\biggr)\,,
\end{align}
where $ u={g}/{16\pi^2}$. This model  is critically equivalent to the nonlinear $\sigma$~-~model, see for a
review~\cite{ZinnJustin:1989mi,Vasilev:2004yr}.
The latter describes a system of two interacting fields~--~basic field $\varphi$ and "auxiliary" field
$\sigma$~--~
with the action~\footnote{Going from~\eqref{Sphi4} to \eqref{NSM} one gets an additional term $\sim \sigma^2$ which, however
is IR irrelevant and can be omitted in the critical regime. }
\begin{align}\label{NSM}
S(\varphi,\sigma)= \int d^d x \left(\frac12(\partial \varphi)^2-\frac12 \sigma \varphi^2\right)\,.
\end{align}
The partition function is given by the path integral
\begin{align}\label{ZJ}
Z(J)= \mathcal{N}^{-1} \int D\varphi\, D\sigma \exp\left\{-S(\varphi,\sigma)+ J_\varphi \varphi +J_\sigma\sigma \right\}.
\end{align}
The $1/N$ expansion for this model is constructed as follows~\cite{Vasiliev:1975mq,Vasilev:2004yr}. One represents the
action~\eqref{NSM} in the form
\begin{align}
S= \int \left(\frac12(\partial \varphi)^2
+\frac12  \sigma K \sigma
-\frac12 \sigma \varphi^2 - \frac12 \sigma K \sigma\right) = S_0+ S_{\text{int}}\,,
\end{align}
where $\int \sigma K \sigma= \int d^d x \int d^dy \, \sigma(x)K(x-y)\sigma(y)$, etc. Thus the kernel $K$ is an inverse
propagator of the $\sigma$ field. It is fixed  by the condition that the term $\sigma K\sigma$ in $S_{int}$ cancels  the LO
$\varphi$ loop insertions to the $\sigma$ lines. Namely,
\begin{align}
K(x)+\frac{N}2 D^2_\varphi(x)=0\,,
\end{align}
where $D_\varphi(x)$ is the propagator of the basic field $\varphi$
\begin{align}
D_\varphi(x)=\frac{a(1)}{4\pi^\mu} \frac{1}{(x^2)^{\mu-1}} && \text{and}
                                  && a(x)=\frac{\Gamma(\mu-x)}{\Gamma(x)}.
\end{align}
Since $D_\sigma =K^{-1}\sim 1/N$ one gets a systematic $1/N$ expansion for \eqref{ZJ}. However, despite the fact that one considers the
theory in non-integer dimensions the loop diagrams are divergent and the theory has to be regularized. The most convenient way
to do it is to modify the kernel $K$ in the free part ($S_0$) of the action \cite{Vasiliev:1975mq},
\begin{align}
K(x)\mapsto K_\Delta(x)= C(\Delta) (M^2 x^2)^{-\Delta} K(x)\,.
\end{align}
The function $ C(\Delta)$ is arbitrary except that it has to satisfy the condition $ C(0)=1$. Different choices of
$ C(\Delta)$ result in  a finite renormalization for  Green functions but do not affect the critical exponents. We fix the
function $C(\Delta)$ by the requirement that the $\sigma$ field propagator takes the form
\begin{align}
D_{\sigma}(x)= \frac1N B(\mu) \frac{M^{2\Delta}}{(x^2)^{2-\Delta}}\,, &&B(\mu)=-\frac{32 a(2-\mu)}{a(2)a^2(1)}\,.
\end{align}
%
The divergences in diagrams arise as poles in $\Delta$ and are removed by the $R$ operation. From now on we will assume the MS scheme,
i.e. $Z$ factors are series in $1/\Delta$, $Z=1+\sum_{k\geq 1} c_k(1/N)/\Delta^k$. The renormalized action takes the
form~\footnote{ Note, that in so-called exceptional dimensions, $d_s=2s/(s-1)$, $s=3,4\ldots$ there are additional
divergences which require the counterterms of the form $(\varphi^2)^s$. In particular, for $d_s=3$ there is  the counterterm
 $(\varphi^2)^3 $, see~\cite{Vasilev:2004yr}. }~\cite{Vasiliev:1975mq}
\begin{align}
S_R(\varphi,\sigma)= \int \left( \frac12 Z_1(\partial \varphi)^2
+\frac12 M^{-2\Delta} \sigma K_\Delta \sigma
-\frac12 Z_2\sigma \varphi^2 - \frac12 \sigma K \sigma\right).
\end{align}
Note, however, that the renormalization is not multiplicative, i.e.
$S_R(\varphi,\sigma)\neq S(\varphi_0,\sigma_0)$. It means that the knowledge of renormalization factors is not sufficient for
determining  critical exponents~\cite{Vasiliev:1975mq,Vasiliev:1993ux}. Nevertheless it was shown in~\cite{Derkachov:1997ch}
that to the
$1/N^2$ accuracy the anomalous dimensions can be expressed via the corresponding renormalization factors in a simple way. The
recipe is the following:  we rescale the propagator of $\sigma$ field by a parameter $u$,
\begin{align}
D_{\sigma}(x)\to D_{\sigma}(x,u)=u\times \frac 1 N B(\mu) \frac{M^{2\Delta}}{(x^2)^{2-\Delta}}\,.
\end{align}
Then the contribution of each diagram, $G$, to the renormalization constant comes with the factor~$u^{n_G}$, where $n_G$  is
the number of
$\sigma$-lines in the diagram. Let $Z$ be the renormalization factor for an operator $\mathcal{O}$,
$[\mathcal{O}](\Phi)=Z\,O_B(\Phi_0)$, $\Phi=\{\varphi,\sigma\}$. In the MS scheme it takes the form
\begin{align}
Z=1 + \frac{1}{\Delta} Z_1(u) + \frac{1}{\Delta^2}Z_2(u) +\ldots,
\end{align}
where $Z_k(u)=\sum_j z_{kj}(u)/N^j$. Then, to the order $1/N^2$ the anomalous dimension of the operator $\mathcal{O}$ can be
obtained as~\cite{Derkachov:1997ch}
\begin{align}\label{eq:gammaZ}
\gamma_{\mathcal{O}} =   2u\partial_u Z_1(u)\Big|_{u=1} + O(1/N^3)\,.
\end{align}
For more details see~\cite{Derkachov:1997ch,Derkachov:1998js}. In certain situations conventional techniques   of
self-consistency equations~\cite{Vasiliev:1981yc,Vasiliev:1981dg} and conformal bootstrap~\cite{Vasiliev:1982dc} are, of course, more
effective. However, the approach outlined above is very convenient for analysis of composite operators, especially with a
nontrivial mixing pattern.

\section{Higher-spin operators}\label{sect:HSO}

\begin{figure}[t]
\centerline{\includegraphics[width=0.49\linewidth]{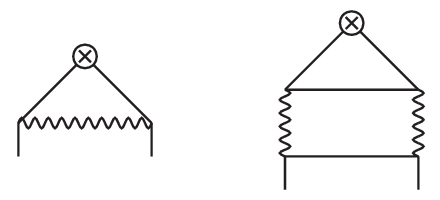}}
\caption{\sf Leading-order  diagrams contributing to the anomalous dimension, $\gamma(s)$. The left diagram contributes
to the non-singlet anomalous dimension $\gamma_{\text{ns}}$ and the right one --  to the pure singlet, $\Delta\gamma$.
Solid lines stand for the propagator of the basic field $\varphi$ and wavy lines for the $\sigma$-field propagator.
\label{fig:diagLO}
}
\end{figure}

We are interested in the critical dimensions of the higher-spin (traceless and symmetric) singlet operators
\begin{align}
J_{\mu_1,\ldots,\mu_s}= \sum_a \varphi^a \partial_{\mu_1}\ldots\partial_{\mu_s} \varphi^a - \text{traces}.
\end{align}
%
In what follows we will not display Lorentz indices explicitly  and adopt  a shorthand notation for the operator,
$J_s\equiv J_{\mu_1,\ldots,\mu_s} $. The operator $J_s$ mixes under renormalization with operators that are total derivatives.
However, since the mixing   has a  triangular form it is irrelevant for  calculation of the anomalous dimensions and can be
neglected. Thus the renormalized operator takes the form
\begin{align}
  [J_s] =  Z(s) J_s\,.
\end{align}
The leading order diagrams contributing to the renormalization factor are shown in Fig.~\ref{fig:diagLO}. The left diagram on
this figure is the only one contributing at this order to the renormalization  of the non-singlet operator. The right
diagram with a closed
$\varphi$ line  cycle, contributes to the singlet operator only. With this in mind we write the answer for the anomalous dimension
of the singlet operator in the form
\begin{align}
  \gamma (s) = \eta + \gamma_{\rm ns}(s) + \Delta \gamma(s).
\end{align}
Here the index $\eta$ determines the anomalous dimension of the field $\varphi$, $\eta=2\gamma_\varphi$, $\gamma_{\rm ns}(s)$
is the anomalous dimension of the non-singlet operator, and
$\Delta \gamma(s)$ is the contribution due to diagrams with a closed $\varphi$-line cycle.
 All contributions except $\Delta \gamma(s)$ are known to the NLO accuracy. The first two expansion coefficients for the index
           $\eta = \eta_1/N + \eta_2/N^2 + O(1/N^3)$
take the form \cite{Vasiliev:1981dg}
\begin{align}
  \eta_1 & = \frac{4(2-\mu) \Gamma(2\mu-2)}{\Gamma(\mu-1)^2 \Gamma(2-\mu)
            \Gamma(\mu+1) } , \notag \\[2mm]
  \eta_2  & = \eta_1^2 \left( - \frac{2\mu^2-3\mu+2}{2-\mu} R(\mu) -3 -\frac3{(\mu-2)^2} -\frac{11}{2(\mu-2)}
  +\frac1{2(\mu-1)} +\frac1{2\mu}\right)\,,
\end{align}
where
\begin{align}
R(\mu)=\psi(1)+\psi(\mu-1)-\psi(2-\mu)-\psi(2\mu-2).
\end{align}
The LO anomalous dimension of the $\sigma$ field ($\gamma_\sigma= \gamma_{\sigma,1}/N+\ldots$) is  \cite{Vasiliev:1981dg}
\begin{align}
\gamma_{\sigma,1}=-2\eta_1\frac{(\mu-1)(2\mu-1)}{2-\mu}\,.
\end{align}
The non-singlet anomalous dimension $\gamma_{\text{ns}}(s)$ has been calculated in~\cite{Derkachov:1997ch} at the order $1/N^2$.
The first two coefficients of the $1/N$ expansion
\begin{align}\label{nsdef}
\gamma_{\text{ns}}(s)=\frac{\eta_1}N \gamma_{\text{ns},1}(s)+ \left(\frac{\eta_1}{N}\right)^2 \gamma_{\text{ns},2}(s)+\ldots
\end{align}
read~\footnote{ In Ref.~\cite{Derkachov:1997ch} the anomalous dimensions of the non-singlet operators symmetric in $O(N)$
indices were calculated. Such operators exist for even spin only. The expression~\eqref{nonsinglet}  is valid for all
$s$. The only difference with the result of~\cite{Derkachov:1997ch}  is an additional sign factor in front of the term  $R(n,\mu)$. }
\begin{align}\label{nonsinglet}
  \gamma_{\text{ns},1}(s)  & = - \frac{\mu (\mu-1)}{j_s(j_s-1)}, \notag\\
  \gamma_{\text{ns},2}(s)  & =  \gamma_{\text{ns},1}(s)\Biggl\{-\frac12\left(\frac1{j_s}+\frac{1}{j_{s}-1}\right)
  \Big(1+\gamma_{\text{ns},1}(s)\Big)
  +\frac12\frac{\mu^2-\mu+1}{\mu(\mu-1)}\gamma_{\text{ns},1}(s)
               +\frac12 \mu(\mu-1)R_s(\mu)
               \notag\\
&\quad -\frac{2(\mu-1)(2\mu-1)}{\mu-2} S(j_s) 
               +\frac{2\mu^2-3\mu+2}{\mu-2} R(\mu) + \frac{\mu^3-4\mu^2+2\mu+2}{(\mu-1)(\mu-2)^2}\Biggr\}\,,
\end{align}
where we introduced the notation, $j_s = s+\mu-1$,  for the  canonical conformal spin. The functions $S(j)$ and $R_s(\mu)$ are
defined as
\begin{align}
  S(j) &=\psi(j)-\psi(\mu-1)\,, \notag\\
  R_s(\mu) &= \int_0^1 d \alpha \int _0^1 d\beta \bar \alpha^{\mu-3}\,\bar \beta^{\mu-3}\, (1-\alpha-\beta)^s =
  R_s^+(\mu)+(-1)^s R_s^-(\mu)\,,
\end{align}
where
\begin{align}
R_-(s)  =\frac{\Gamma^2(\mu-2) s!}{\Gamma(s+2\mu-3)},
&&
R_+(s) 
=\frac1{s+1}\int_0^1d\alpha\,\alpha^{s+\mu-2} \,{}_2F_1 \genfrac{(}{|}{0pt}{}{1,3-\mu}{s+2}\alpha\biggl)\,.
\end{align}

Singlet operators exist only for even spins, so that from now on we assume that $s$ is even. At $1/N$ order only one
diagram -- the rightmost diagram in Fig.~\ref{fig:diagLO} -- contributes to the pure singlet anomalous dimension
\begin{align}
\Delta \gamma(s) & =\frac{\eta_1}N \Delta\gamma_1(s)+\left(\frac{\eta_1}N\right)^2\Delta\gamma_2(s)+O(1/N^3)\,.
\end{align}
Calculating this diagram and using Eq.~\eqref{eq:gammaZ} we reproduce the known result~\cite{Lang:1991kp}
\begin{align}
\Delta  \gamma_1(s)
  = -
  \frac{2\mu(\mu-1)\Gamma(2\mu-2)\Gamma(s+1)}{j_s(j_s-1)\Gamma(s+2\mu-3)}
=  2\gamma_{\text{ns},1}(s)\frac{\Gamma(2\mu-2)\Gamma(s+1)}{\Gamma(s+2\mu-3)}\,.
\end{align}
Thus to the leading order $1/N$ the singlet  anomalous dimension is
\begin{align}\label{singlet1N}
  \gamma(s) = \frac{\eta_1}N \left( 1 -\frac{\mu (\mu-1)}{j_s(j_s-1)} \left(1+
  \frac{2\Gamma(2\mu-2)\Gamma(s+1)}{\Gamma(s+2\mu-3)} \right) \right).
\end{align}
Note, that for $s=2$ the anomalous dimension vanishes as it should be since the spin two current corresponds to the
energy-momentum tensor. We also remark that the spin dependence of the LO singlet anomalous dimensions~\eqref{singlet1N}  is
exactly the same as in the Gross-Neveu model, see e.g.~\cite{Muta:1976js,Manashov:2016uam}.


\subsection{Singlet current at $1/N^2$}
\begin{figure}[t]
\centerline{\includegraphics[width=0.99\linewidth]{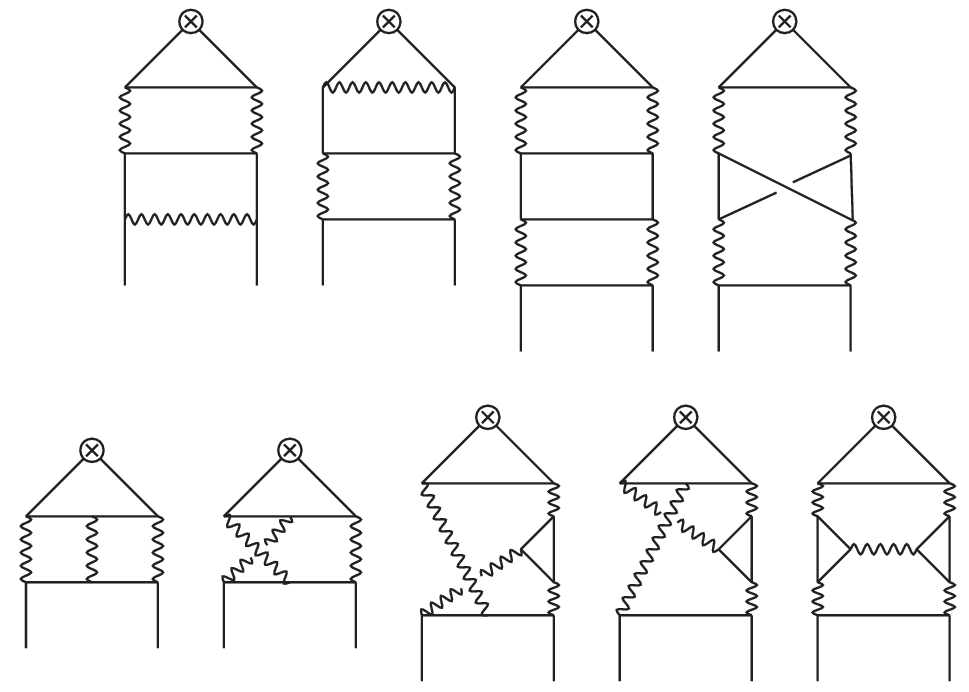}}
\caption{\sf Feynman diagrams of different topologies, $D_1,\ldots, D_9$, contributing to the
$1/N^2$ anomalous dimension of the singlet current $J_{s}$.
\label{fig:diagNLO}
}
\end{figure}

The diagrams which contribute to the pure singlet anomalous dimension at the order $1/N^2$ can be split in two groups. The first
one comprises the self-energy and vertex corrections to the leading order diagram (eight different diagrams in total). The
diagrams from the second group are shown in Fig.~\ref{fig:diagNLO}. The diagrams from the first group can be effectively
calculated with the help of technique developed in~\cite{Derkachov:1997ch}. We give some details of this calculation in
Appendix \ref{app:vse}.

Next, the first three diagrams in the Fig.~\ref{fig:diagNLO} are easy to calculate. All other diagrams  have only a
superficial divergency. Since we are interested only in a residue at the $\Delta$ pole   the regulator $\Delta$ can be removed
from the
$\sigma$-lines and placed  on one of the
$\varphi$-lines. For $\Delta=0$ the basic $\sigma\varphi^2$ vertex has the property of uniqueness and can be transformed with
the help of the star-triangle relation
\begin{figure}[H]
\centerline{\includegraphics[width=0.63\linewidth]{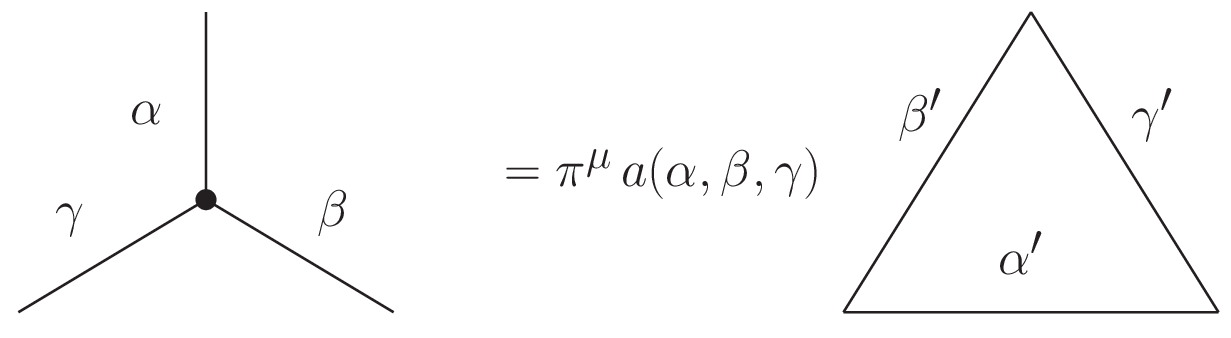}}
\label{fig:Star}
\end{figure}
\noindent which holds if $\alpha+\beta+\gamma=2\mu$. Here $a(\alpha,\beta,\gamma)\equiv a(\alpha)a(\beta)a(\gamma)$ and
$\alpha'=\mu-\alpha$, etc. Using the standard technique~\cite{Vasilev:2004yr} one can find rather straightforwardly
the contribution of each diagram to the  renormalization constant of the singlet current. We collected answers for individual
diagrams in Appendix~\ref{app:diagrams}.

Before presenting the answer for the singlet anomalous dimensions let us note that if one  is interested only in the $d=3$ result
the calculation of the last three diagrams can be greatly simplified. It should be stressed here that we are talking about the
pole part of the diagrams only. Using the star~-~triangle relation one derives in $d=3$:
\begin{figure}[H]
\centerline{\includegraphics[width=0.89\linewidth]{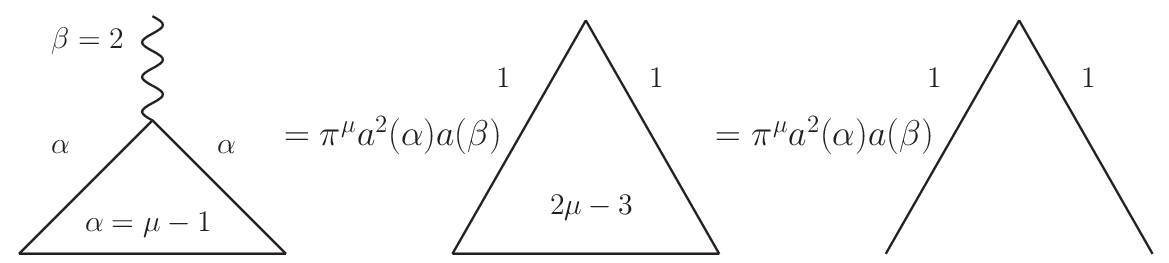}}
\label{fig:Trick}
\end{figure}
\noindent Since $d=2\mu=3$ the horizontal line on the rightmost diagram disappears. In this way it is easy to check that the
contributions of the third and fourth diagrams in the second line in Fig.~\ref{fig:diagNLO} vanish in $d=3$, while the
last diagram is reduced to a simple ladder-type diagram by application of the chain rule.

Collecting all terms, our answer for the NLO singlet anomalous dimension takes the form:
\begin{align}\label{N-expansion}
\gamma(s) &=
\frac{\eta_1}N \gamma_{1}(s)+ \left(\frac{\eta_1}{N}\right)^2 \gamma_{2}(s)+\ldots
\notag\\
&=\frac{\eta_1}N \Big(1+\gamma_{\text{ns,1}}(s)+\Delta\gamma_{1}(s)\Big)
-\frac12\left(\frac{\eta_1}N\right)^2\Big(1+\gamma_{\text{ns,1}}(s)+\Delta\gamma_{1}(s)\Big)
 \biggl[\gamma_{\text{ns,1}}(s) \left(\frac1{j_s}+\frac1{j_s-1}\right)
\notag\\
&\quad
+
\Delta\gamma_{1}(s) \left(\frac1{j_s}+\frac1{j_s-1}+\psi(j_s+\mu-2)-\psi(j_s+2-\mu) -2+\psi(3-\mu)-\psi(\mu-1)\right)\biggr]
\notag\\
&\quad +\frac{\eta_2}{N^2} + \left(\frac{\eta_1}{N}\right)^2\Biggl\{
\gamma_{\text{ns,1}}(s)\Biggl[
\frac12\frac{\mu^2-\mu+1}{\mu(\mu-1)}\gamma_{\text{ns,1}}(s) +\frac12\mu(\mu-1) R_s(\mu)
\notag\\
&\quad
-\frac{2(\mu-1)(2\mu-1)}{\mu-2} S(j_s)+\frac{2\mu^2-3\mu+2}{\mu-2}\,R(\mu) +\frac{\mu^3-4\mu^2+2\mu+2}{(\mu-1)(\mu-2)^2} \Biggr]
\notag\\
&\quad
+\Delta\gamma_1(s)\Biggl[ \left(\mu(2-\mu)-\frac2{2-\mu}\right)
S(j_s)
-\frac12\Delta\gamma_1(s)-\frac1{2j_s(j_s-1)}
\notag\\[2mm]
&\quad
+\frac{2(\mu-1)(2\mu-1)}{\mu-2}\Psi(j_s) + \frac{2}{\mu-2}\ R(\mu)
-\frac{\mu^4-4\mu^3 +9\mu^2-6\mu-2}{(\mu-1)(\mu-2)^2}
\notag\\[2mm]
&\quad
+
 \frac{\Big(\frac12\Delta\gamma_1(s) +4(2\mu-3)\gamma_{\text{ns,1}}(s)\Big)j_s(j_s-1) +
 \mu(\mu-1)(j_s+\mu-3)(j_s+2-\mu)}{(2- \mu)(j_s+1-\mu)(j_s+\mu-2)}\Phi(j_s)
\notag\\
&\quad +\frac{2\mu(\mu-1)(2\mu-3)}{(2-\mu)^2}\Biggl(
-\frac{2\mu-3}{s(s+2\mu-3)}\left(\Psi(j_s)+\frac1{s+2\mu-3}+\frac1{2-\mu}-R(\mu)\right)
\notag\\
&\quad +\frac{\Gamma(2\mu-2)\Gamma(s)}{\Gamma(s+2\mu-2)}\biggl(
S(j_s)-\psi(s+1)+\psi(1)-\frac1{2\mu-3}+\frac1{s+2\mu-3}+\frac1{2-\mu}-R(\mu)
\biggr)
\notag\\[2mm]
&\quad
+\frac{\Gamma(2\mu-2)}{\Gamma(\mu-2)\Gamma(s+\mu-1)}\sum_{m=0}^{s-1} C^{s-1}_m
\frac{\Gamma(s-m) \Gamma(\mu-1+m)\Gamma(s+\mu-2-m)}{(m+1)^2\Gamma(s+2\mu-3-m)}
 \Biggl)
 \Biggr]
\Biggr\}.
\end{align}
Here
\begin{align}
\Psi(j) &=\psi(j+\mu-2)+\psi(j+2-\mu)-2\psi(j) -\psi(1)-\psi(2\mu-2)+2\psi(\mu-1)\,,
\notag\\
\Phi(j)&=\Big[\psi(j+\mu-2)+\psi(j+2-\mu)-2\psi(j)+\psi(\mu-1)-\psi(1)
 -  \mathrm{J}(j,\mu)\Big]
\end{align}
and the function $\mathrm{J}(j,\mu)$ is defined 
by
\begin{align}
\mathrm{J}(j,\mu) &=
\frac{\Gamma(j)}{\Gamma(\mu-2)s! } \int_0^1 d\alpha \alpha^{2\mu-4+s}\int_0^1 d\beta  \frac{\beta^{\mu-2}\bar\beta^{s}}{1-\alpha\beta}
=\frac{\mu-2}{j(j+\mu-2)} {}_3F_2 \genfrac{(}{|}{0pt}{}{1,\mu-1,j+\mu-2}{j+\mu-1,j+1} 1 \biggl)
\,.
\end{align}
The expression~\eqref{N-expansion} passes several consistency checks. First, it can be verified that for $s=2$ the
singlet anomalous dimension vanishes,  $\gamma(s=2)=0$. We remark also that the non-singlet spin one  current is conserved  and,
hence, its anomalous dimension vanishes, $\eta+\gamma_{\text{ns}}(s=1)=0$.

Second, the large spin asymptotic of $\gamma(s)$ complies  with  the CFT prediction~\cite{Alday:2015eya,Alday:2015ewa}. It was
noticed in~\cite{Dokshitzer:2005bf,Basso:2006nk} that if one represents  anomalous dimensions of higher-spin operators in the
form
\begin{align}\label{gamma-f}
\gamma(s)=f\left(j_s+\frac12\gamma(s)\right)
\end{align}
then the asymptotic expansion of the  function $f(j)$ has  a rather specific form. Namely, it is given by the sum of terms
\begin{align}\label{as-exp}
 \left(j-\frac12\right)^{-\Delta_q}\sum_{k\geq 0} \frac{a_{q,k}}{(j(j-1))^k}\,.
\end{align}
Excluding the prefactor,   this series is invariant under $j\to 1-j$ save that the coefficients $a_{q,k}$ are allowed to be
functions (polynomials) of
$\ln (j-1/2)$. For more detail see Refs.~\cite{Alday:2015ewa,Alday:2015eya}. In the perturbative expansion Eq.~\eqref{gamma-f} takes the form
\begin{align}\label{g-f-e}
\gamma(s)=f_1(j_s)+\frac12 f_1(j_s) f'_1(j_s) + f_2(j_s)+\ldots
\end{align}
Comparing it with \eqref{N-expansion} one finds that $f_1(j_s)$ is the LO anomalous dimension, while the second term in
\eqref{g-f-e} is contained in the first two lines in \eqref{N-expansion}. Thus the large spin expansion of all other terms in
\eqref{N-expansion} starting from the third line has to have the form ~\eqref{as-exp}. The asymptotic expansion of all
contributions, except the diagram $D_9$~\eqref{D9},  can be easily calculated and has  the form~\eqref{as-exp}. For the
diagram $D_9$ we checked this property in $d=3$ only.

Third, the anomalous dimension of the higher-spin currents in $\varphi^4$ model in $4-2\epsilon$ expansion are known with
four-loop accuracy \cite{Derkachov:1997pf}. Restoring $O(N)$ factors for individual diagrams given
in~\cite{Derkachov:1997pf} we obtain for
$\gamma(s)$
\begin{align}\label{u-expansion}
\gamma(s) & = \frac{N+2}3\Biggl\{
u^2 \frac{(s-2)(s+3)}{6s(s+1)}  - u^3\frac{(N+8)}{9s(s+1)}\left[2S_1(s)+\frac{s^4 + 2s^3 - 39s^2 - 16s + 12}{8s(s+1)}\right]
\notag\\
&\quad u^4 \frac5{864}(-N^2+18N+100)-\frac{u^4}{s(s+1)}\Biggl[\frac{N+2}{18}\left(S_1(s)-\frac{11s^4+20s^3+15s^2-6s-6}{2s^2(s+1)^2} \right)
\notag\\
&\quad
+\frac{N^2+6N+20}{27}\left(S_2(s)+S_1^2(s) - S_1(s)\frac{4s^2+2s-1}{s(s+1)} +\frac{8s^4-4s^3-13s^2-s+3}{4s^2(s+1)^2} \right)
\notag\\
&\quad
+\frac{5N+22}{27}\biggl(2S_2(s)+S_1^2(s)-\frac{s^2+s-4}{s(s+1)}K_2(s)
\notag\\
&\quad
-S_1(s)\frac{11s^2+7s-2}{s(s+1)}
+\frac{42s^4+52s^3+3s^2-7s+3}{2s^2(s+1)^2}
 \biggr)\Biggr]
 \Biggr\} + O(u^5)\,,
\end{align}
where $S_k(n)=\sum_{m=1}^n 1/m^k$ and $K_2(n)=\sum_{m=1}^{n} (-1)^{m+1}/m^2$. Expanding \eqref{N-expansion} in $\epsilon$ for
$\mu=2-\epsilon$ and \eqref{u-expansion} in $1/N$ for $u=u_*$, Eq.~\eqref{critical-u}, we find  complete agreement between
both results.

\begin{figure}[t]
\centerline{\includegraphics[width=0.42\linewidth]{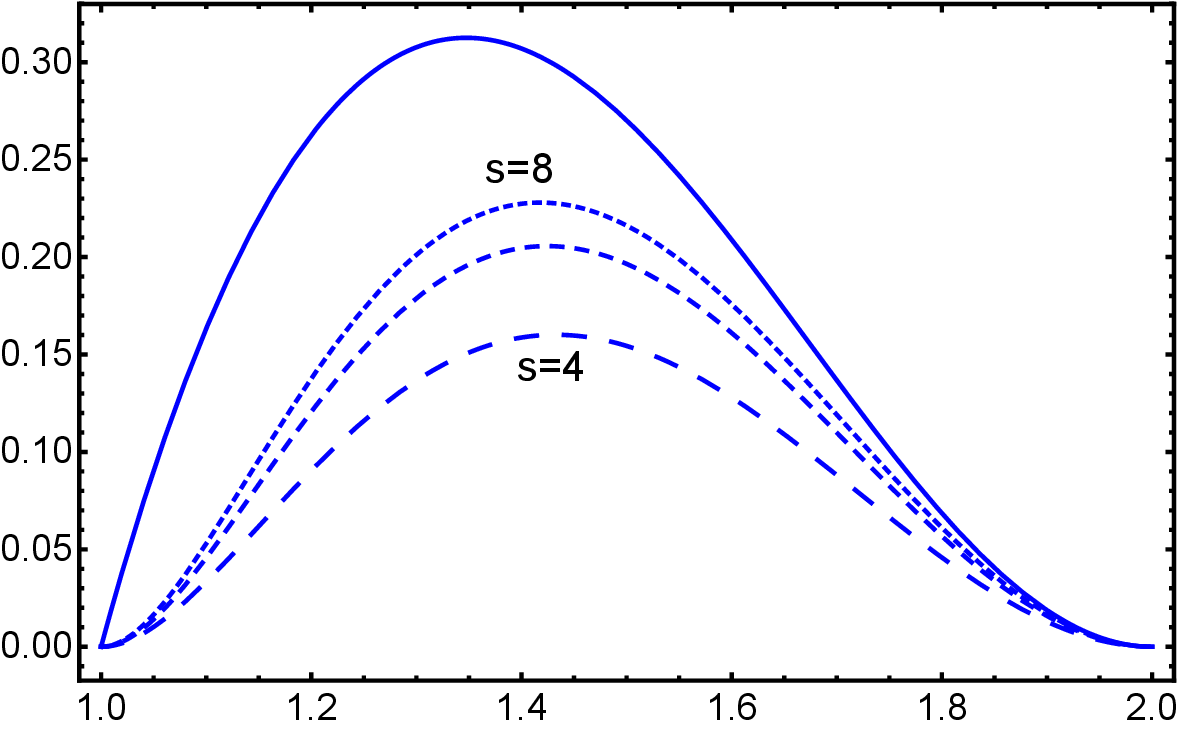}\hspace{10mm}\includegraphics[width=0.42\linewidth]{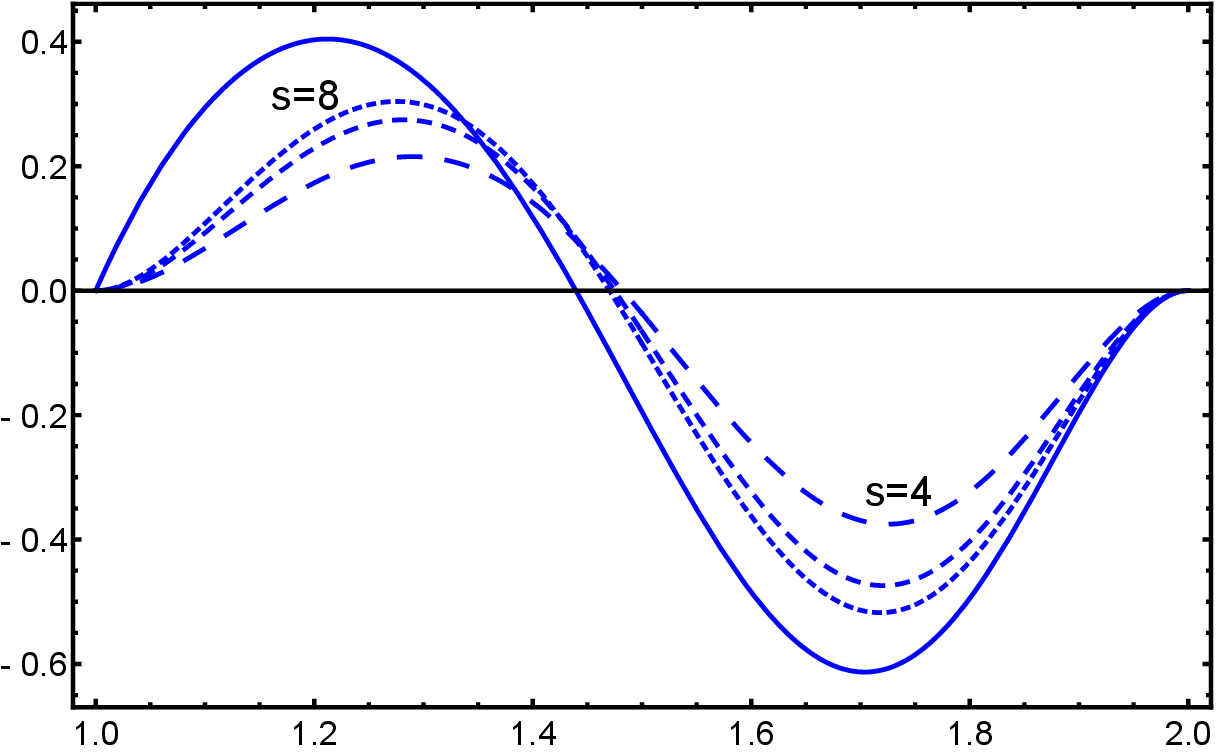}}
\caption{\sf Anomalous dimensions of the singlet currents as function of $\mu$ for $s=4,6,8$. The order-$1/N$ correction is on the left plot and the order-$1/N^2$ correction is on the right. The solid line corresponds to the limiting value of $\eta_1$ (on the left) and to $\eta_2$ (on the right).
\label{fig:plo468}
}
\end{figure}

The anomalous dimensions of the singlet currents as a function of dimension $d=2\mu$ for few lower spins are shown in
Fig.~\ref{fig:plo468}. The LO anomalous dimensions are positive in the whole interval $2<d<4$ while the NLO correction change
the sign near $d=3$. It explains a relative smallness of  NLO corrections in $d=3$.

\section{$d=3$ reduction and higher-spin masses}\label{sect:d3}
\begin{figure}[t]
\centerline{\includegraphics[width=0.5\linewidth]{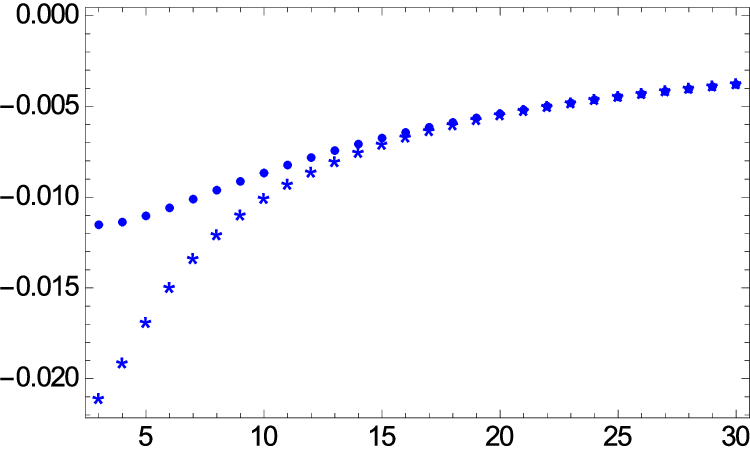}}
\caption{\sf $\gamma(4)-\eta$ as a function of $N$ for $N=3,...,30$. Stars correspond to $1/N$ expansion and dots to $4-2\epsilon$ expansion.}
\label{fig:plotcmp}
\end{figure}

In three dimensions the results can be considerably simplified. First of all,
\begin{align}
\eta_1&=\frac{8}{3 \pi ^2}\,, & \eta&=\frac{\eta_1}{N}\left(1-\frac{8}{3N}\eta_1+\ldots\right)\,.
\end{align}
After some simplifications we obtain for the non-singlet anomalous dimensions in three dimensions
\begin{align*}
 \gamma_{\text{ns},1}(s)&=-\frac{3}{\left(4 s^2-1\right)}\,,
 \notag\\
    \gamma_{\text{ns},2}(s)&=\frac{3}{4 \left(4 s^2-1\right)}\Biggr\{-\frac{128 s^2}{9}-6 \pi  (-1)^s s+\frac{11}{2 s-1}-\frac{6}{(2 s-1)^2}-\frac{3}{2 s+1}+\frac{6}{(2 s+1)^2}+\frac{158}{9}\\
    &\qquad\qquad\qquad\qquad
    -32 \log (2)+6 s S_1\left(\frac{s}{2}-\frac{1}{4}\right)-6 sS_1\left(\frac{s}{2}
    -\frac{3}{4}\right)-16 S_1\left(s-\frac{1}{2}\right) \Biggr\}.
\end{align*}
For the anomalous dimensions of the singlet currents we have
\begin{align*}
\gamma_1(s) &=\frac{2(s-2)}{2s-1}\,,
\notag\\
  \gamma_2(s)
  &=  \frac{3}{4s^2-1} \bigg(-\frac{32 s^2}{9}-\frac{\left(13 s^2+14 s+6\right) \log (2)}{s}-\frac{3}{2} \pi  s+\frac{3}{2} s \left(S_1\left(\frac{s}{2}+\frac{3}{4}\right)-S_1\left(\frac{s}{2}+\frac{1}{4}\right)\right)\\
  &\quad+\frac{3 (-1 - s + s^2)}{s}\left(S_1\left(\frac{s}{2}\right)-S_1\left(\frac{s+1}{2}\right)-S_1(s+1)\right)-\frac{(s+2) (7 s+6)}{2 s}S_1\left(s-\frac{1}{2}\right)\\
  &\quad+\frac{\left(13 s^2+3 s+3\right)}{s}S_1(s)+13 s-\frac{9}{s+1}+\frac{1}{2 s-1}-\frac{6}{(2 s-1)^2}
  -\frac{3}{2 s+1}+\frac{9}{2 s+3}-\frac{9}{s}+\frac{152}{9}\bigg).
\end{align*}
%
It may be interesting to compare the results of the large-$N$ expansion with the perturbative results in $4-2\epsilon$ dimensions for $\epsilon=\tfrac12$, which is displayed on Fig.~\ref{fig:plotcmp} for the $s=4$ current. We see that as $N$ increases the two approximation converge to each other.

\begin{figure}[t]
\centerline{\includegraphics[width=0.5\linewidth]{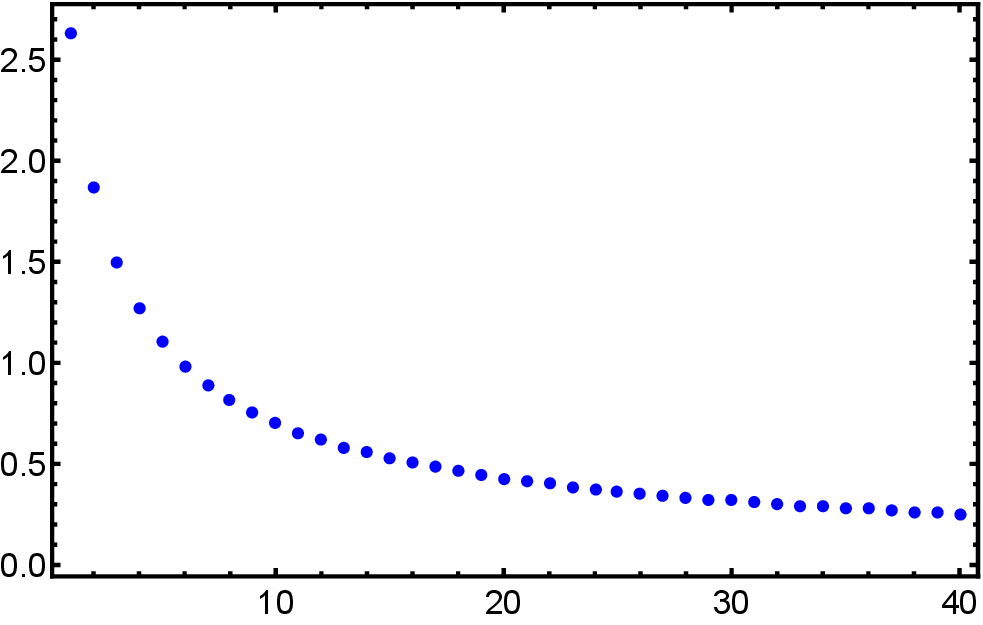}}
\caption{\sf Function $\varkappa(2k)$.
}\label{fig:plomass}
\end{figure}

Now  we can write down the effective masses of higher-spin fields in $AdS_4$:
\begin{align}
\begin{aligned}
    \delta m_s^2=&\frac{2}{N}(s-2)\eta_1 +\frac{\eta_1^2}{N^2}\frac{3 }{2 s+1}\bigg\{-\frac{\left(13 s^2+14 s+6\right) \log (2)}{s}+\frac{\left(13 s^2+3 s+3\right) }{s}S_1(s)\\
    &-\frac{(s+2) (7 s+6)}{2 s} S_1\left(s-\frac{1}{2}\right)-\frac{32 s^2}{9}+\frac{41 s}{3}-\frac{9}{s+1}-\frac{3}{2 s+1}+\frac{9}{2 s+3}-\frac{9}{s}\\
    &+\frac{137}{9}-\frac{3}{2}  \pi  s+\frac{3 \left(s^2-s-1\right)}{s}\left(S_1\left(\frac{s}{2}\right)
    -S_1\left(\frac{s+1}{2}\right)-S_1(s+1)\right)\\
    &+\frac{3}{2} s \left(S_1\left(\frac{s}{2}+\frac{3}{4}\right)-S_1\left(\frac{s}{2}+\frac{1}{4}\right)\right)\bigg\}.
\end{aligned}
\end{align}
The order-$1/N$ correction is linear, while the effective mass up to the order-$1/N^2$ can be written as
\begin{align}
\delta m_s^2&=2\eta(s-2)\big(1+\eta\varkappa(s)+\ldots\big).
\end{align}
At large spin $\varkappa(s)=\tfrac{39}{8}\frac{\log s}{s}+...$  and we also plot $\varkappa(s)$ in Fig.~\ref{fig:plomass}.

\section{Summary}\label{sect:summary}

We have calculated the $1/N^2$ corrections to the anomalous dimensions of the singlet higher-spin currents in the $O(N)$
vector model. Also, using the results of Ref.~\cite{Derkachov:1997pf} we recovered the four-loop  anomalous dimensions in the $O(N)$
model and checked that the $1/N$ and $\epsilon$ expansions for the anomalous dimensions  are in complete agreement with each other.

The $1/N^2$ expression for the anomalous dimensions~\eqref{N-expansion} is rather involved but simplifies considerably in three dimensions. We have also related them to the two-loop radiative corrections to the masses of higher-spin fields.

It has been known that the LO critical dimensions of the singlet higher-spin currents coincide in the $O(N)$ and Gross-Neveu models with some identification of the expansion parameters. Our result  shows that it is no longer true at the NLO order even in $d=3$. This also implies that the NLO anomalous dimensions of the higher-spin currents in Chern-Simons matter theories have a more complicated form than the one observed at the LO.

\section*{Acknowledgments}
\label{sec:Aknowledgements}
We are grateful to Johan Henriksson for spotting two typos in the formulas. 
This work was supported by Deutsche Forschungsgemeinschaft (DFG) with the grants 
$\text{MO~1801/1-2}$ (A.M.) and SFB/TRR 55 (M.S.)
%
%
and  in part  by the Russian Science Foundation grant 14-42-00047 in association with Lebedev Physical Institute and by the
DFG Transregional Collaborative Research Centre TRR 33 and the DFG cluster of excellence ``Origin and Structure of the
Universe" (E.Sk.).

\appendix

\section*{Appendices}
\addcontentsline{toc}{section}{Appendices}

\renewcommand{\theequation}{\Alph{section}.\arabic{equation}}
\renewcommand{\thesection}{{\Alph{section}}}
\renewcommand{\thetable}{\Alph{table}}
\setcounter{section}{0} \setcounter{table}{0}


\section{Numerical Values}\label{app:diagrams}
We collect below numerical values of the order-$1/N^2$ anomalous dimensions of the singlet currents. It is worth stressing
that
$\eta_1=\frac{8}{3 \pi ^2}$ is the same for the Gross-Neveu and for the critical vector models. It is convenient to give
anomalous dimensions as multiples of $(\eta_1)^2$, see Eq.~\eqref{nsdef}.  Conservation of the $O(N)$~-~current implies
$\gamma_{\text{ns}}(1)=0$ and we obtain for a few lowest spins
\begin{align}
\begin{aligned}
\gamma_{\text{ns},2}(2)&= -\frac{696}{125}
\approx
-5.568\,,
& \gamma_{\text{ns},2}(3)&= -\frac{263104}{128625}
\approx -2.04551\,,
\\
\gamma_{\text{ns},2}(4)&=  -\frac{548936}{138915} \approx -3.9516\,,
& \gamma_{\text{ns},2}(5)&= -\frac{8406592}{3773385}\approx -2.22786\,.
\end{aligned}
\end{align}
Conservation of the stress-tensor implies $\gamma(2)=0$ and we have
\begin{align}
\begin{aligned}
\gamma_2(4)&= -\frac{1544}{3087} \approx -0.500162\,,  & \gamma_2(6)&= -\frac{233008}{259545}\approx -0.897756\,,  \\
\gamma_2(8)&= -\frac{22279496}{19144125}  \approx -1.16378\,,  & \gamma_2(10)&=-\frac{248880040436}{183833374725} \approx-1.35383\,.
\end{aligned}
\end{align}
Let us note that the numerical values for the singlet currents are by an order of magnitude smaller than those in the
Gross-Neveu model. Let us also write down the singlet anomalous dimensions in a way that makes it clear that the second order
corrections are relatively  small
\begin{align}
\begin{aligned}
\gamma_2(4)&=\frac{4}{7}\frac{\eta_1}{N} \left(1-0.2364\frac{1}{N}+\ldots\right),  &
\gamma_2(6)&=\frac{8}{11}\frac{\eta_1}{N} \left(1-0.3335\frac{1}{N}+\ldots\right),
\\ \gamma_2(8)&=\frac{4}{5}\frac{\eta_1}{N} \left(1-0.3930\frac{1}{N}+\ldots\right),  &
\gamma_2(10)&=\frac{16}{19}\frac{\eta_1}{N} \left(1-0.4343\frac{1}{N}+\ldots\right).
\end{aligned}\end{align}

\section{Diagrams}\label{app:diagrams} 
In this appendix we collect the results for the diagrams shown in Fig.~\ref{fig:diagNLO}. The expressions below give
the divergent part of  diagrams with subtracted counterterms. The symmetry factors are already included in these  expressions.
We obtained for the diagrams $D_k$, $k=1,\ldots,9$
{\allowdisplaybreaks\begin{align}
D_1& = \frac{\eta_1^2}4 \gamma_{\text{ns},1}(s) \Delta\gamma_1(s) \,
\Biggl(-\frac1{6\Delta^2} +\frac1{3\Delta}\biggl[-1+\frac1{\mu+s-1}+\frac1{\mu+s-2}
\notag\\
&\quad  \hspace{40mm}+\psi(2\mu-3+s)+\psi(3-\mu)-\psi(\mu-2)-\psi(2)\biggr]\Biggr)\,,
\notag\\[1mm]
D_2 & =\frac{\eta_1^2}{12}\gamma_{\text{ns},1}(s) \Delta\gamma_1(s) \,
\Biggl(-\frac1{\Delta^2} +\frac1{\Delta}\biggl[-1+\frac1{\mu+s-1}+\frac1{\mu+s-2}-\frac1{\mu-2}+\psi(2)-\psi(s+1)\biggr]
\Biggr)\,,
\notag\\[1mm]
D_3 & =\frac{\eta_1^2}{32} \Delta\gamma^2_1(s) \,\Biggl(-\frac1{\Delta^2}+\frac2\Delta\biggl[
-1+\frac1{\mu+s-1}+\frac1{\mu+s-2}
\notag\\
&\quad
\hspace{40mm} +\psi(2\mu+s-3)-\psi(s+1)-\psi(\mu-1)+\psi(3-\mu)
\biggr] \Biggr)\,,
\notag\\
D_4 & =
-\frac1\Delta\,\frac{\eta_1^2}{16}\,\Delta\gamma^2_1(s)\, \frac{(s+\mu-1)(s+\mu-2)}{(2- \mu) s(s+2\mu-3)}\,\Phi(j_s)
\,,
\notag\\[2mm]
D_5 & =-\frac1{6\Delta} \eta^2_1\, \Delta \gamma_1(s)\,\frac{\mu(\mu-1)(3-\mu)}{\mu-2}\biggl(\psi(\mu+s-1)-\psi(\mu-2)\biggr)\,,
\notag\\[2mm]
D_6 & = -\frac1{6\Delta} \eta_1^2\Delta\gamma_1(s)\,\frac{\mu(\mu-1)}{2-\mu}\Biggl(
\frac1{2-\mu} + \frac{(s+1)(s+2\mu-4)}{s(s+2\mu-3)}\times \Phi(j_s)
\Biggr),
\notag\\[2mm]
D_7 & = D_8 
=-\frac{\eta_1^2}{4\Delta} \gamma_{\text{ns},1}(s) \Delta\gamma_1(s) \frac{(2\mu-3)(s+\mu-1)(s+\mu-2)}{(2- \mu)s(2+2\mu-3)}\,\Phi(j_s)\,.
\end{align}
For the last diagram we get
\begin{align}\label{D9}
D_9&= -\frac{\eta_1^2}{5\Delta} \Delta\gamma_1(s)\frac{\mu(\mu-1)(2\mu-3)}{(2-\mu)^2}\Biggl\{
-\frac{2\mu-3}{s(s+2\mu-3)}\left(\Psi(j_s)+\frac1{s+2\mu-3}+\frac1{2-\mu}-R(\mu)\right)
\notag\\
&\quad +\frac{\Gamma(2\mu-2)\Gamma(s)}{\Gamma(s+2\mu-2)}\biggl(
S(j_s)-\psi(s+1)+\psi(1)-\frac1{2\mu-3}+\frac1{s+2\mu-3}+\frac1{2-\mu}-R(\mu)
\biggr)
\notag\\[2mm]
&\quad
+\frac{\Gamma(2\mu-2)}{\Gamma(\mu-2)\Gamma(s+\mu-1)}\sum_{m=0}^{s-1} C^{s-1}_m
\frac{\Gamma(s-m) \Gamma(\mu-1+m)\Gamma(s+\mu-2-m)}{(m+1)^2\Gamma(s+2\mu-3-m)}
 \Biggl\}\,.
\end{align}}%

\section{Vertex and self-energy corrections}\label{app:vse}

In this Appendix we discuss the calculation of the self-energy and vertex corrections diagrams. In total there are eight
different diagrams which arise from the SE and vertex corrections to the LO pure singlet diagram. The calculation of SE
diagrams is rather straightforward but cumbersome while the vertex corrections could be rather involved. The reason for this
is that the diagrams with vertex corrections contain, evidently, divergent subgraph and, therefore, one cannot remove the
regulator $\Delta$ from the $\sigma$ lines and use the uniqueness property of the $\sigma \varphi^2$ vertex. However it is
helpful to take into account that the model under consideration is a conformal one. The form of two- and three- point
correlators in CFT is fixed up to normalization factors by the scaling dimensions of the fields. Namely, the dressed (full)
propagators and 1PI irreducible three point function $\Gamma_{\sigma\varphi\varphi}$ have the form
\begin{align}
D_\varphi(x)={\widehat A}/{x^{2\Delta_\varphi}}\,, &&D_\sigma(x)={\widehat B}/{x^{2\Delta_\sigma}}\,,
\end{align}
and
\begin{align}
\gamma_R(z,x,y)\equiv \Gamma_{\sigma\varphi\varphi}(z,x,y)= \widehat{Z} (z-x)^{-2\alpha} (z-y)^{-2\alpha} (x-y)^{-2\beta}\,.
\end{align}
Here
\begin{align}
\Delta_\varphi  &=\mu-1+\gamma_\varphi\,, & \Delta_\sigma&=2+\gamma_\sigma\,,
\notag\\
\alpha& =\mu-1-\gamma_\sigma/2\,, &\beta &=2-\gamma_\varphi+\gamma_\sigma/2\,.
\end{align}
The explicit expressions for the factors $\widehat{A},~\widehat{B},~\widehat{Z}$ at the order-$1/N$  can be found in
Ref.~\cite{Derkachov:1997ch}. One can use this information in order to avoid a tedious calculation of the individual diagrams.
Namely, it was shown in~\cite{Derkachov:1997ch} that  the contribution to the anomalous dimension at the
$1/N^2$ order  due to the SE and vertex corrections to the $1/N$ diagram  can be extracted from the   same  diagram
 with the dressed propagators and vertices.

 Let us consider  a logarithmically divergent diagram.  For such a diagram  the number of $\varphi$ lines is equal
 to the number of basic vertices and twice a number of the $\sigma$ lines,
 \begin{align}\label{NNV}
 N_\varphi = N_V=2N_\sigma\,.
 \end{align}
 Replacing
 propagators and vertices by full propagators and vertices one gets a superficially divergent diagram. It has to be regularized
 by introducing the regulator $\Delta$ in any line. The resulting  diagram has  a simple pole in $\Delta$
 \begin{align}
G=\frac1{\Delta} R + F\,.
 \end{align}
The contribution to the anomalous dimension which comes from SE and vertex corrections diagram is equal to
\begin{align}
\delta\gamma^{SE+V} =-{2r^{(1)}}/{N^2}\,,
\end{align}
where $r^{(1)}$ comes from the expansion of the residue $R$ in $1/N$,  $R=r^{(0)}/N+ r^{(1)}/N^2+O(1/N^3)$.

\begin{figure}[t]
\centerline{\includegraphics[width=0.95\linewidth]{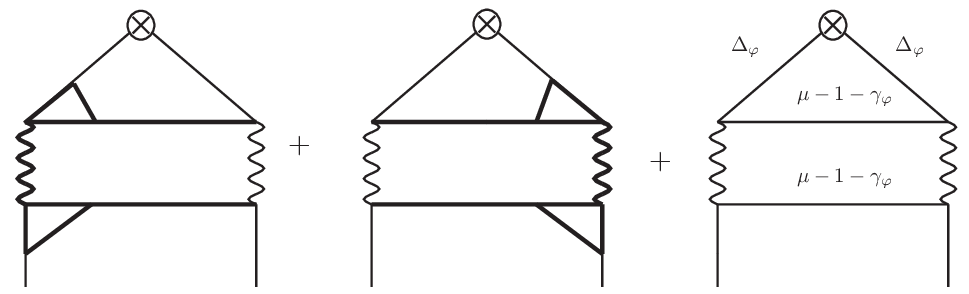}}
\caption{\sf Vertex and Self-Energy correction diagrams for the pure singlet diagram.
\label{fig:SEV}
}
\end{figure}

 The triple vertices in the modified diagram has the uniqueness property that usually simplifies calculation greatly.
Moreover, it is not necessary to replace all vertices  and propagators at once. The contributions from lines and vertices  are
additive~\cite{Derkachov:1997ch}. One can replace a subset of  lines and vertices, $S_1\subset S$, satisfying the
condition~\eqref{NNV}, calculate the corresponding diagram $G^{(1)}$ and find the coefficient $r^{(1)}_1$. Then the same can be done for the next subset,
$S_2$, and so on. If sets $S_k$ are not intersecting, $ \bigcap_k S_k=\emptyset$ and $\bigcup_k S_k=S$, then
$\delta\gamma= -\frac{2}{N^2}\sum_k r_k^{(1)}$. If the sets  $S/\bigcup_k S_k=S_+$ and $\bigcap_k S_k=S_-$ are not empty then we
have to add the contributions from the elements in $S_+$ and subtract those in $S_-$. We illustrate this rule on the example
of the pure singlet diagram, Fig.~\ref{fig:diagLO}. The corresponding decomposition is shown in Fig.~\ref{fig:SEV}.

In the leftmost diagram we replaced  two left vertices, the left $\sigma$ line and  two horizontal $\varphi$ lines. In the
middle diagram --  two right vertices, the right $\sigma$ line and the two horizontal $\varphi$ lines.   So  the contribution of
the horizontal lines is counted twice. Thus we have to add the contribution from the lines attached to the operator vertex and
subtract contribution from the horizontal lines. It is done by adding the rightmost diagram. All the diagrams are superficially
divergent and have to be regularized by shifting index of one of the $\sigma-$lines by $\Delta$. All these diagrams (first two
are obviously equal each other) can be easily calculated with the help of a chain integration rule and the star-triangle
relation.

After simple calculation we find for the residue $r^{(1)}$ of a simple pole of the diagrams $D_{10}$, $D_{11}$ and  $D_{12}$:
\begin{align}
r^{(1)}_{D_{10}} &=r^{(1)}_{D_{11}} = -\frac12 \eta_1 \Delta\gamma_1(s)\Biggl\{\chi_1\left( \psi(\mu+s-1)-\psi(\mu-2)-\frac{\mu}{\mu-1}\right)
\notag\\
&\quad
+\frac12\gamma_{\sigma,1} \left(\psi(s+1)+ \psi(s+2\mu-3)-2\psi(1)+\psi(3-\mu)-\psi(\mu-1)+6\frac{3-\mu}{\mu-2}\right)
\notag\\
&\quad
+ \eta_1\left[ \frac{2\mu^2-3\mu+2}{\mu-2}\cdot R(\mu) + 2(\mu-1)\left(4-\frac 7{(\mu-2)^2}\right)\right]\Biggr\},
\end{align}
where $\chi=-\eta-\gamma_\sigma=\chi_1/N + O(1/N^2)$ and
\begin{align}
r^{(1)}_{D_{12}} &= -\frac1{4}\eta_1^2 \Delta\gamma_1(s) \Biggl\{ 2\Big(\psi(s+\mu-2)-\psi(\mu-1)+1\Big)
\notag\\
&\quad
+\psi(\mu-1)-\psi(3-\mu)-\psi(s+2\mu-3)+\psi(s+1)
\Biggr\}.
\end{align}
%


\providecommand{\href}[2]{#2}\begingroup\raggedright\endgroup

\end{document}